\input mont.sty
\input dc_mont.sty
\input epsf.sty
\hsize=16cm \vsize=21cm
\hfuzz=0.2cm
\tolerance=400
\def\msbar{$\overline{\hbox{MS}}$}

\parindent6mm

\hyphenation{Leut-wyler}
\rightline{FTUAM--00-18; %hep-ph/xxxxxxxx
}

\noindent{\twelverm  QCD Calculations of Heavy Quarkonium States}
{\footnote*{\petit Talk given at the conference `` Hyperons, Charm and Beauty Hadrons", 
Valencia, June 2000. To be published in Nucl. Phys. Suppl.}}

\phantom{Typeset with \physmatex}
\vskip0.4cm
\noindent F. J. Yndur\'ain
\vskip0.3cm
\noindent Departamento de F\'\i sica Te\'orica, C-XI,
Universidad Aut\'onoma de Madrid,\hb
Canto Blanco, 28049-Madrid, Spain. E-mail: fjy@delta.ft.uam.es
\vskip0.6cm
\noindent  Recent results on the QCD 
analysis of bound states of heavy $\bar{q}q$ quarks are reviewed,  
paying attention to what can be derived from the theory with a reasonable degree of rigour. 
We report a calculation of $\bar{b}c$ bound states;  
a very precise evaluation 
of $b,\,c$ quark masses from quarkonium spectrum; the NNLO evaluation of 
$\Upsilonv\to e^+e^-$; and a discussion of 
power corrections. For the $b$ quark {\sl pole} mass we get, including 
$O(m_c^2/m_b^2)$ and
 $O(\alpha_s^5\log \alpha_s)$ corrections, $m_b=5.020\pm0.058\,\gev$; 
and for the $\overline{MS}$ mass the result, correct to  $O(\alpha_s^3)$, $O(m_c^2/m_b^2)$,  
$\bar{m}_b(\bar{m}_b)=4.286\pm0.036\,\gev$. 
For the decay $\Upsilonv\to e^+e^-$, higher corrections are too large 
to permit a reliable calculation, but we can 
predict a toponium width of $13\pm1\,\kev$.

\begindc{
\noindent{\fib 1. INTRODUCTION}
\smallskip
\noindent In the present note we are going to review some aspects of the QCD analysis
 of heavy quarkonia, $\bar{t}t$, $\bar{c}c$, $b\bar{c}$ and especially $\bar{b}b$ states.  
 The validity
 of {\sl ab initio} QCD calculations varies from one 
case to another. For the energy levels of toponium with $n\leq5$, 
$n$ being the principal quantum number, and for the energy 
of the ground state in bottomium, we have a very 
satisfactory situation, almost comparable to that in positronium calculations. 
For bottomium hyperfine splitting, and 
for the energy of the {\sl charmonium} ground 
state, the situation is less favourable, although 
reasonably under control. Less favourable still 
is what one has  for the 
energy levels of 
bottomium with $n=2$. Here NLO and NNLO perturbative and  
the leading nonperturbative corrections are comparable to 
 the LO results. 
The same is true for the decay $\Upsilonv\to e^+e^-$, and 
even for toponium decay, $T\to e^+e^-$, higher corrections are a bit too large for comfort.  
For higher states,  for the decays of $n=2$ bottomium or for the 
ground state of charmonium, a {\sl rigorous} QCD 
evaluation is out of the question. 
There are methods that have been devised in the literature to deal 
with this, essentially more or less justified models; 
we do not discuss them now. 
The interested reader may find information, and 
 references, in e.g. my Lisbon 
lectures\ref{1}. 

This does not mean that even in the favourable cases all questions are settled, 
and in particular
 we discuss here  power-like corrections, {\sl at short distances}, 
where there is still some controversy.  
\smallskip
\phantom{BRRR}
\noindent{\fib 2. NONRELATIVISTIC QCD POTENTIAL, AND CORRECTIONS}
\smallskip
\noindent To analyze the lowest states of heavy quarks one proceeds as follows: 
first, in the nonrelativistic (NR) approximation one finds the 
potential that 
governs the dynamics. 
This potential is evaluated to increasing orders of 
perturbation theory; 
at present it is known to one and two\ref{2} loops. 
One can thus write
$$H=H^{(0)}+H_{\log};\equn{(2.1a)}$$
where $H^{(0)}$ may be solved exactly and contains the coulombic  
part of the interaction. For $\bar{q}q$ states,
$$H^{(0)}=2m+
\dfrac{-1}{m}\lap-\dfrac{C_F\tilde{\alpha}_s(\mu^2)}{r},\equn{(2.1b)}$$
$$\eqalign{\tilde{\alpha}_s(\mu^2)=&
\alpha_s(\mu^2)
\left\{1+c^{(1)}\dfrac{\alpha_s(\mu^2)}{\pi}+c^{(2)}\dfrac{\alpha_s^2}{\pi^2}\right\};\cr
c^{(1)}=&a_1+\dfrac{\gammae\beta_0}{2}\cr
c^{(2)}=&\gammae\left(a_1\beta_0+\dfrac{\beta_1}{8}\right)
+\left(\dfrac{\pi^2}{12}+\gammae^2\right)\dfrac{\beta_0}{4}+a_2.\cr}$$
The expression for the $a_i$ may be found in 
e.g. ref.~1, $m$ is the pole mass of the quark, and $H_{\log}$ is
$$\eqalign{H_{\log}=&
\dfrac{-C_F\beta_0\alpha_s(\mu^2)^2}{2\pi}\, \dfrac{\log r\mu}{r}
+\dfrac{-C_F\beta_0^2\alpha_s^3}{4\pi^2}\,\dfrac{\log^2 r\mu}{r}\cr
+&\dfrac{-C_F\alpha_s^3}{\pi^2}\,
\left(a_1\beta_0+\dfrac{\beta_1}{8}+\dfrac{\gammae\beta_0^2}{2}\right)\dfrac{\log r\mu}{r}.\cr
}$$

One solves exactly the coulombic Schr\"odinger equation,
$$H^{(0)}\Psiv_{nl}=E^{(0)}_{nl}\Psiv_{nl} ,\equn{(2.2)}$$
treating the terms in $H_{\log}$ as perturbations. 
 Relativistic corrections are then added, 
to be also treated as perturbations. These relativistic corrections 
include spin-dependent ones, known at tree level 
since a long time (they are essentially like for positronium) and to one loop 
from refs.~3,4.
Moreover, and because of the nonabelian character of QCD, 
 we have mixed relativistic--one loop corrections for 
the spin-independent piece, first 
evaluated in ref.~5 using the method of equivalent potentials of 
Gupta et al. (see e.g. 
ref.~4). Concentrating on the spin-independent 
hamiltonian we then write
$$H_{\rm SI}^{\rm e.p.}=H^{(0)}+H_{\log}+\dfrac{-1}{4m^3}\lap^2+\dfrac{C_F\alpha_s}{m^2r}\lap
+\dfrac{C_Fb_1\alpha_s^2}{2mr^2} \equn{(2.3)}$$
and $b_1=\tfrac{1}{2}C_F-2C_A$. Here the superindex ``e.p." means that 
the hamiltonian was obtained with the 
method of equivalent potentials. 

The hamiltonian has been re-evaluated recently by 
Brambilla et al.\ref{6}; these authors 
generalize the nonrelativistic (heavy quark) effective theory (NRQCD\ref{7}), 
matching  then the  Green's functions to a potential description 
(pNRQCD\ref{8}). In this way they find,
$$\eqalign{H_{\rm SI}^{\rm pHQET}=H^{(0)}+H_{\log}+
\dfrac{-1}{4m^3}\lap^2\cr
+\dfrac{\pi C_F\alpha_s}{m^2}\delta({\bf r})+
\dfrac{C_F\alpha_s}{2r^3m^2}{\bf L}^2-\dfrac{C_F\alpha_s}{2m^2}\{\dfrac{1}{r},\lap\}\cr
+\dfrac{-C_FC_A\alpha_s^2}{2mr^2}\cr.} \equn{(2.4)}$$

This is {\sl different} from (2.3). 
The difference is due to the fact that in the derivation of 
(2.3) one works with the S-matrix, 
while for (2.4) one uses Green's functions. The difference 
between (2.3) and (2.4) 
vanishes when taking expectation values between coulombic 
wave functions, say, solutions of (2.2): so they will produce the same 
energy spectrum, at least to order $\alpha_s^4$. 
At higher orders, (2.3) and (2.4) would produce different results; 
but also new terms will appear in the two formalisms: 
their equivalence, or non-equivalence, has not 
been proved. For numerical results to $O(\alpha_s^4)$ 
for all states see ref.~9; here we present those for the ground state energy. 
We write 
$$E^{\rm p.t.}_{10}=2m-m\dfrac{C_F^2\widetilde{\alpha}_s^2}{4}+
\delta_{\rm p.t.}E_{10}.
\equn{(2.5a)}$$
The label ``p.t." in $E^{\rm p.t.}_{nl}$ indicates that we have as yet 
only used results deduced from perturbation theory; the full expression would be
$$E_{nl}=E^{\rm p.t.}_{nl}+\delta_{\rm NP}E_{nl}.$$
$\delta_{\rm NP}E_{nl}$ embodies the nonperturbative contributions, 
to be discussed below.  
The $\delta_{\rm p.t.}E_{10}$ is, with $a=2/mC_F\widetilde{\alpha}_s$,
$$\eqalign{\delta_{\rm p.t.}E_{10}=&-\dfrac{10}{8m^3\,a^4}
-\dfrac{3C_F\alpha_s}{m^2a^3}+\dfrac{C_Fb_1 \alpha_s^2}{ma^2}\cr
&-\dfrac{C_Fc_2^{(L)}\alpha_s^3}{\pi^2a}\;
\left[\log\dfrac{a\mu}{2}+\psi(2)\right]\cr
&-\dfrac{C_F\beta_0^2\alpha_s^3}{4\pi^2a}\,
\Big\{\log^2\dfrac{a\mu}{2}-\dfrac{\gammae}{2}\log\dfrac{a\mu}{2}\cr
+&\dfrac{3+3\gammae^2-\pi^2+6\zeta_3}{12}\Big\},
\cr}\equn{(2.5b)}$$
$c_2^{(L)}=a_1\beta_0+\tfrac{1}{8}\beta_1+\tfrac{1}{2}\gammae\beta_0^2.$

The (nominally) leading, $\log\alpha_s$ corrections of next order are also known. 
They include a logarithmic correction to the static potential\ref{10}, 
and a relativistic one-loop correction\ref{5,11}. With the full result (as given in ref.~6) 
one has, for $\mu=2/a$, 
$$\delta_{[\alpha_s^5\log\alpha_s]} E_{10}=
-m[C_F+\tfrac{3}{2}C_A]C^4_F\alpha_s^5(\log\alpha_s)/\pi.\equn{(2.5c)}$$
In the calculations we will include, for the $b$ 
quark case,  a correction\ref{11} of order $m_c^2/m_b^2$,
$$\delta_{[m_c^2/m_b^2]} E_{10}=2m_b\dfrac{3T_F\alpha_s}{2\pi}\,\dfrac{m_c^2}{m_b^2}.
\equn{(2.5d)}$$ 
We can invert (2.5) to obtain a precise 
determination of $c,\,b$ quark masses from those of the $J/\psi,\,\Upsilonv$ 
particles. The determination includes 
nonperturbative effects, to be discussed below. 
As input parameters we take the recent determination\ref{12},
$$\eqalign{\Lambdav(n_f=4,\,\hbox{three loops})=0.283\pm0.035\;\gev\cr
\left[\;\alpha_s(M_Z^2)\simeq0.117\pm0.024\;\right],\cr}
$$ 
and for the gluon condensate the value
$\langle\alpha_sG^2\rangle=0.06\pm0.02\;{\gev}^4.$ 

Then comes the matter of the renormalization point, $\mu$. 
As \equs~(2.5b,c) show, logarithms are avoided until order $\alpha_s^5$ 
if choosing $\mu=\mu_0=2/a$, and this will be our central value here.
Then, for the $b$ quark, we find
$$\eqalign{m_b=&
5.020\pm0.043\,(\Lambdav)\;\mp0.005\,(\langle\alpha_sG^2\rangle)\;\cr
&^{-0.031}_{+0.037} \;(\hbox{vary}\; \mu^2\;{\rm by}\,25\%)
\cr\pm& 0.006\pm0.017\;({\rm other\; th.\;uncert.})\cr
=&5.020\pm0.058\,\gev.\cr}
\equn{(2.6a)}$$
To obtain this, we have included the correction of \equn{(2.5c)}, and 
for the ``theoretical error" also that estimated in 
ref.~13 for the perturbative evaluations. The value 
of the \msbar\ mass  
that corresponds to this is, taking into account 
$O(\alpha_s^3)$ and leading $O(m_c^2/m_b^2)$ corrections\ref{14},
$$\bar{m}_b(\bar{m}_b)=4272\pm43\;\mev.\equn{(2.6b)}$$
We present a summary of QCD calculations of the quark masses, with 
increasing accuracy.
\bigskip
\setbox0=\vbox{
\setbox1=\vbox{\offinterlineskip\hrule
\halign{
&\vrule#&\strut\hfil#\hfil&\vrule#&\strut\quad#\quad&\vrule#&\strut\quad#\quad&\vrule#&\strut\quad#\quad&\vrule#&\strut\quad#&#\cr
 height2mm&\omit&&\omit&&\omit&&\omit&&\omit&\cr 
&\kern0.2em Ref.\kern0.2em&&$m_b({\rm pole})$&&$\bar{m}_b(\bar{m}_b^2)$&&$m_c({\rm pole})$& \cr
 height1mm&\omit&&\omit&&\omit&&\omit&\cr
\noalign{\hrule} 
height1mm&\omit&&\omit&&\omit&&\omit&\cr
&TY&& $4971\pm72$&&$4401^{+21}_{-35}$\vphantom{$4^{4^4}_{4_4}$}&&
$1585\pm 20\,^*$&\cr
&PY&& $5065\pm60$&&$4455^{+45}_{-29}$\vphantom{$4^{4^4}_{4_4}$}&&
$1866^{+215}_{-133}$&\cr
&Here&& $5022\pm58$&&$4272\pm43$&&
$-$&\cr
 height1mm&\omit&&\omit&&\omit&&\omit&\cr
\noalign{\hrule}}
\vskip.05cm}
\centerline{\box1}
{\petit
\centerline{ $b$,  $c$ quark masses. $(^*)$
 Systematic errors not included.}}
\vskip-0.2cm
\centerrule{0.2cm}
\smallskip
\setbox2=\vbox{\hsize=0.95\hsize 
\petit{
\noindent
TY: Titard and Yndur\'ain\ref{5}. $O(\alpha_s^3)$ plus $O(\alpha_s^3)v$, $O(v^2)$ 
for $m$;
$O(\alpha_s^2)$ for $\bar{m}$. 
 Rescaled for 
$\Lambdav(n_f=4)=283\,\mev$.\hb
PY: Pineda and Yndur\'ain\ref{9}. Full $O(\alpha_s^4)$ for $m$; 
$O(\alpha_s^2)$ for $\bar{m}$. Rescaled for 
$\Lambdav(n_f=4)=283\,\mev$.\hb
Here: This calculation. $O(\alpha_s^4)$, $O(\alpha_s m_c^2/m_b^2)$ 
and $O(\alpha_s^5\log\alpha_s)$ 
 for $m$; $O(\alpha_s^3)$ and  $O(\alpha_s^2 m_c^2/m_b^2)$  for  $\bar{m}$. 
Values not given for the \msbar\ $c$ 
quark mass, as the higher order terms are as large as the leading ones.}}
\centerline{\box2}}
\centerline{\box0}
\centerrule{0.3cm}
\smallskip
It is to be noted that pure QCD quarkonium determinations 
of $m_b$ have remained remarkably  
 stable since the earlier evaluations (ref.~5). [The 
variation of the {\sl published} values is due to the variation of the 
favoured central value of $\Lambdav$, from $200$ to $283$ \mev]. 
Also, the results depend very little on the value of $\mu$ chosen. 
Thus, even taking $\mu=m\simeq5\,\gev$ one gets
$$m_b=4901\pm60\,\mev, $$
reasonably close to (2.6a), where we had elected $\mu=2/a\simeq 2.88\,\gev$. 
The estimates of $\bar{m}_b(\bar{m}_b)$ are less stable, doubtlessly 
because of the large size of the $O(\alpha_s^3)$ corrections; 
but it is interesting that this more precise spectroscopic determination of 
\msbar\ mass of the $b$ quark mass given here agrees with 
recent determinations, based on sum rules, 
also accurate to $O(\alpha_s^3)$. These  give\ref{15}
$$\bar{m}_b(\bar{m}_b)=4260\pm100\,\mev.$$
We will discuss \msbar\ masses further later on. 

The hyperfine splitting 
can likewise be evaluated, getting a prediction for the $\eta_b$ mass:
$$M(\Upsilonv)-M(\eta_b)=53.3\pm12.5\;\mev.$$

The methods discussed for quarkonium can be extended with some work to 
hybrid states like $b\bar{c}$. 
The advantage here is that, since the masses of $b,\,c$ quarks can 
be expressed in terms of those of the $\Upsilonv, \, J/\psi$ 
particles, the theoretical errors are smaller than what one would get for 
charmonium. A recent $O(\alpha_s^4)$ calculation\ref{16} gives
$$M_{10}(b\bar{c})=6323\pm10\pm20\;\mev,$$
the first error being perturbative, 
the second coming from estimated nonperturbative effects.
\smallskip
\noindent{\fib 3. POWER CORRECTIONS.  THE NONPERTURBATIVE VACUUM;
 RENORMALONS;\hb SATURATION}
\smallskip
\noindent As stated above, a calculation such as that in (2.6) includes leading 
nonperturbative effects. 
These are obtained by realizing that the motion of the 
$\bar{q}q$ pair takes place in the physical vacuum, chock full 
of soft gluons and light quark pairs so that, in particular, 
we have a nonzero value for the gluon condensate:
$$\langle{\rm vac}|:\alpha_s G^2(0):|{\rm vac}\rangle\equiv
\langle\alpha_sG^2\rangle\neq 0.$$
The effects of this were first considered by Leutwyler and Voloshin\ref{17} 
(see also refs..~18,~5)
 and amount to a shift for (say) the ground state energy of quarkonium of
$$\delta_{\rm NP}^{\langle\alpha_sG^2\rangle}=
m\dfrac{\pi\epsilon_{10}\langle \alpha_s:G^2:\rangle}{(mC_F\alpha_s)^4},\equn{(3.1)}$$
$\epsilon_{10}\simeq 1.5$. 
This is  zero to all orders of 
perturbation theory, because so is the gluon condensate. 
On dimensional grounds one expects $\langle\alpha_sG^2\rangle\sim \Lambdav^4$, 
so that (3.1) is of order $\Lambdav^4/m^3\alpha_s^4$.

This is not the only power correction that may appear. 
We may have {\sl renormalons}. 
Let us consider a one-gluon exchange diagram, for $\bar{q}q$ 
scattering. If we dress the gluon propagator with loops then the 
corresponding potential, in momentum space, is
$$\tilde{V}(k)=\dfrac{-4\pi C_F}{k^2}\,\dfrac{4\pi}{\beta_0\log(k^2/\Lambdav^2)},
\equn{(3.2)}$$
and we have substituted the one-loop expression for $\alpha_s(k^2)$. 
 (3.2) is undefined for {\sl soft} gluons, with
 $k^2\simeq\Lambdav^2$. As follows from the general theory of 
singular functions, the ambiguity is of the form $c\delta(k^2-\Lambdav^2)$: upon 
Fourier transformation this produces 
an ambiguity in the $x$-space potential of 
$\delta V(r)=c[\sin \Lambdav r]/r$. 
At short distances we  expand this in powers of $r$ and 
find\ref{19,20}
$$\delta_{\rm renorm.} V(r)\sim C_0+C_1r^2+\dots\,.\equn{(3.3)}$$
The same result may be obtained with the more traditional method of Borel transforms.
 This coincides with the short distance behaviour of the nonperturbative potential 
 as determined by Dosch and Simonov\ref{18}.

The situation just described applies for states $\bar{q}q$ at 
short distances; but not so short that zero frequency gluons cannot separate 
the pair. If this last is the case, soft gluons do 
not resolve the $\bar{q}q$ pair and only {\sl see} 
a dipole. 
The generated renormalon may then be seen (ref.~20)
 to correspond to the contribution of the 
gluon condensate in the Leutwyler-Voloshin mechanism.

What happens to the Aglietti and Ligeti renormalon? 
In fact, we have other long-distance power corrections. 
It is clear that the pole mass is defined purely 
in perturbation theory, and indeed one can check that a renormalon 
ambiguity appears already at one loop\ref{21}. 
Likewise, the coulombic potential is defined 
so that it vanishes at infinity: but, for 
confined quarks, ``infinity" is equivalent to the 
confinement radius, $R\sim 1/\Lambdav$. 
Actually, mass and potential 
renormalons cancel for the constant and quadratic terms 
in (3.2), in the very short distance regime; this has been 
verified by a detailed calculation in ref.~22, 
thus substantiating the intuitive arguments of ref.~20, 
already mentioned.
\medskip
\setbox0=\vbox{\hsize 6.2cm\hfil\epsfxsize 5cm\epsfbox{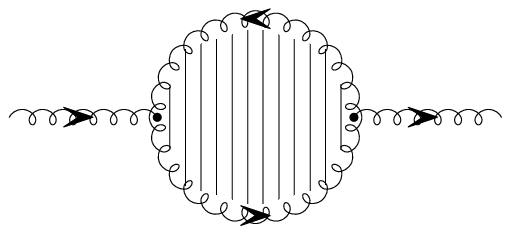}\hfil}
\setbox1=\vbox{\hsize 6cm\captiontype\figurasc{figure 1. }
 {``Filled in" gluon loop.}\hb
\vskip.2cm}
\setbox9=\vbox{\hsize 6.4cm
\centerline{{\box0}}
\centerline{\box1}}
\centerline{{\box9}}

Renormalons and the  mechanism of Leutwyler--Voloshin are not the only possible 
sources of power corrections; we next consider  
 {\sl saturation}. We note that the 
ambiguities we have found  are 
associated with small momenta or,  equivalently,  long 
distances. However,  the {\sl singularities} are clearly 
spurious. Indeed, not only the theory 
should be well defined but, because of 
confinement, long distances are never attained: the theory possesses an internal 
infrared cut-off 
of the order of the confinement radius, $R\sim\Lambdav^{-1}$. To try and 
implement it we  consider the gluon propagator. To one loop 
it gets a correction involving the vacuum polarization tensor.
 Neglecting quarks this is, in $x$-space, 
given by an expression like

$$\eqalign{\Piv^{aa'}_{\alpha\beta}(x,0)\sim& g^2f_{abc}f_{a'de}
\langle0|\int\dd^4y_1\,\dd^4y_2\cr
{\rm T}B_b^\alpha(y_1)\partial_\mu& B_{c\alpha}(y_1)
B_d^\beta(y_2)\partial_\nu B_{e\beta}(y_2)|0\rangle\cr
+&\cdots\,.\cr}$$
We can take into account the {\sl long distance} interactions 
by introducing a string between the field products at finite distances. In 
 matrix notation for the gluonic fields, 
${\cal B}^\mu=t_aB_a^\mu$, this 
is incorporated  by replacing
$$\eqalign{{\cal B}^\alpha(y_1){\cal B}^\beta(y_2)\cr
\to
{\cal B}^\alpha(y_1){\rm P}\left(\exp\,\ii\int^{y_1}_{y_2}\dd z^{\mu}\,{\cal B}_\mu(z)\right)
{\cal B}^\beta(y_2).\cr}$$

The process may be described as ``filling the loop" (see \fig~1) by introducing 
all exchanges between the gluonic lines there. If we furthermore 
replace the perturbative vacuum $|0\rangle$ 
by the nonperturbative one $|{\rm vac}\rangle$, then we get a dressed propagator
$$D^{\mu\nu}_{\rm dressed}(k)=
D^{(0)\mu\nu}(k)\dfrac{4\pi}{\beta_0\log(M^2+k^2)/\Lambdav^2},$$
and  $M^2$ is related to the gluon condensate at 
finite distances, $\langle G(x)G(0)\rangle_{\rm vac}$. (for 
more details and references, see e.g. Simonov's lectures\ref{23}).

This suggests a {\sl saturation} property of the coupling constant at 
small momenta (long distances) so that the expression for the 
running coupling constant should be modified according to
$$\eqalign{\alpha_s(k^2)=&\dfrac{4\pi}{\beta_0\log k^2/\Lambdav^2}\cr
\to&\alpha^{\rm sat}_s(k^2)=\dfrac{4\pi}{\beta_0\log (k^2+M^2)/\Lambdav^2}.\cr}\equn{(3.4)}$$
It is certain that an expression such as (3.4) incorporates, to some extent, 
long distance properties of the QCD interaction. For example, if we take (3.4) 
with $M=\Lambdav$ in the tree level potential for heavy quarks, this 
becomes the {\sl Richardson potential}\ref{23}
$$\eqalign{V^{(0)}({\bf k})=&-\dfrac{4\pi C_F\alpha_s({\bf k}^2)}{{\bf k}^2}\cr
\to&
V^{(0),{\rm sat}}({\bf k})=
-\dfrac{16\pi^2C_F}{\beta_0{\bf k}^2\log (k^2+\Lambdav^2)/\Lambdav^2}.\cr}$$
When one has ${\bf k}^2\gg\Lambdav^2$, the short distance coulombic potential is, of course, 
recovered. For ${\bf k}^2\ll\Lambdav^2$, however,
$$V^{(0),{\rm sat}}({\bf k})\simeqsub_{{\bf k}^2\ll\Lambdav^2}
\dfrac{16\pi^2C_F\Lambdav^2}{\beta_0{\bf k}^4},$$
whose Fourier transform gives
$$V^{(0),{\rm sat}}(r)\simeqsub_{r\gg\Lambdav^{-1}}(\hbox{constant})\times r,$$
i.e., a linear potential. Indeed, a reasonably accurate description 
of spin-independent splittings in quarkonia states is
 obtained with such a potential. However, this long-distance linear potential induced 
by saturation in the Richardson model is 
the {\sl fourth component} of  a Lorentz vector, 
while we know that the Wilson linear potential, as 
obtained, e. g., in the stochastic vacuum model or in lattice calculations, should be 
a Lorentz scalar. This is also obvious on 
phenomenological grounds as 
this potential  has to provide attraction for {\sl quark-quark} states 
in e.g. baryons. It thus follows that  
the linear potential obtained from saturation can be only of limited phenomenological
 use. 

Saturation also gives a linear 
correction to the {\sl short} distance coulombic 
interaction. The possibility of such a correction has 
been discussed by several people; see, for example refs..~20,~23. 
Personally I am not impressed by these arguments. 
It is clear that the QCD perturbative series is {\sl not} convergent; 
in the case of quarkonia, the coefficients are not even analytic 
as one has terms in $\log\alpha_s$ (cf. (2.5c) and \sect~4 below). 
The results one finds are then dependent on how one sums  and, 
moreover, it is not guaranteed that results valid for {\sl large} orders of 
perturbation theory will hold in the real world, where one knows at most three 
nontrivial terms. 
It is the author's belief that only if the summation method 
is rooted on solid physics it is likely to represent an 
improvement; otherwise, estimates of nonperturbative effects 
become pure guesswork. In this respect, the method of taking 
into account the nonperturbative nature of the physical vacuum 
by considering the effect of nonzero values for the correlators 
stands some chance of being meaningful, as indeed phenomenological calculations 
seem to indicate.
\smallskip
\noindent{\fib 4. FURTHER DISCUSSION OF THE 
\msbar\ MASS: A MATTER OF CONVERGENCE}
\smallskip
\noindent Write, for a heavy quark,  
$$\eqalign{\bar{m}(\bar{m})\equiv& m/\{1+\delta_1+\delta_2+\delta_3+\cdots\},\cr
\delta_n=&C_n\alpha_s^n,\cr}\equn{(4.1a)}$$
with the coefficients $C_n$ given in refs.~14 (see also ref.~11). Numerically one has
$$\eqalign{\delta_1(b)=&0.090,\cr
\delta_2(b)=&0.045,\cr
\delta_3(b)=&0.029;\cr}\quad
\eqalign{\delta_1(c)=&0.137,\cr
\delta_2(c)=&0.108,\cr
\delta_3(c)=&0.125.\cr}\equn{(4.2)}$$
Thus, the series relating pole and \msbar\ masses is at the 
edge of the region of convergence for the $b$ 
quark, and clearly diverges for the third order contribution for the $c$ quark.

We can use (4.2) and the formula for $M(\Upsilonv)$ in terms of 
the pole mass to express the former in terms of $\bar{m}_b$:
One finds, for the perturbative contributions, 
 neglecting $m_c^2/m_b^2$, and
with $\Lambdav$ as before,  
$$\eqalign{M(\Upsilonv)=&2\bar{m}_b(\bar{m}_b)
\Big\{1+C_F\dfrac{\alpha_s(\bar{m})}{\pi}\cr
+&7.559\left(\dfrac{\alpha_s(\bar{m})}{\pi}\right)^2
+43.502\left(\dfrac{\alpha_s}{\pi}\right)^3\Big\}.\cr}
\equn{(4.3)}$$
(We have taken $n_f=4$). 
This does not look particularly convergent, and is certainly not 
an  improvement over the perturbative expression using the pole mass where one has,   
for the choice $\alpha_s(\mu=C_Fm_b\alpha_s)$ and 
still neglecting the masses of quarks lighter 
than the $b$, 
$$\eqalign{M(\Upsilonv)=&2m_b\Big\{1-2.193\left(\dfrac{\alpha_s}{\pi}\right)^2-
24.725\left(\dfrac{\alpha_s}{\pi}\right)^3\cr
-&
458.28\left(\dfrac{\alpha_s}{\pi}\right)^4+
897.93\,[\log\alpha_s] \left(\dfrac{\alpha_s}{\pi}\right)^5\Big\}.\cr}\equn{(4.4)}$$
To order three, (4.4) is actually better than (4.3); 
thus, and at least to third order, it is unclear that there is a 
connection between the rate of convergence and
 the use of pole or \msbar\ mass. 
What is more, at least the size of the {\sl known} 
power corrections (namely, those stemming from the 
gluon condensate) do not favour the expression in terms of the \msbar\ mass. 
Indeed, if we evaluate the gluon condensate corrections to 
the direct formula (4.3) for $M(\Upsilonv)$ in terms of 
$\bar{m}_b(\bar{m}_b)$, it turns out that they are    
  larger than than what one would have 
 for the expression in terms of the pole mass, (4.4)  
($\sim 80$ against $\sim9$ \mev), 
because of the definition of the renormalization point.  
\smallskip 
\noindent{\fib 5. DECAYS OF QUARKONIA}
\smallskip
\noindent Besides energies of bound states, 
once one has an effective interaction 
it is possible to evaluate, at least in principle, decay rates such as 
$\Upsilonv\to e^+e^-$. To relative order $\alpha_s$ (NLO) 
this has been done in ref.~5 (see also ref.~9) using the 
method of variations and the evaluation of ref.~25 for the 
hard part, $\delta_{\rm rad}$. 
We give here the exact result of a Rayleigh--Schr\"odinger
 perturbative calculation. One finds,
$$\eqalign{\Gammav(\Upsilonv\rightarrow& e^+e^-)=\Gammav^{(0)}
\left[1+\delta_{\rm wf}+\delta_{\rm rad}+\delta_{\rm NP}\right]^2\,,\cr
\Gammav^{(0)}=&2\left[\dfrac{Q_b\alpha_{\rm QED}}{M(V)}\right]^2
\left(mC_F{\alpha}_s(\mu^2)\right)^3;\cr 
\delta_{\rm wf}+\delta_{\rm rad}=&
\Bigg\{\dfrac{3\beta_0}{4}\left(\log\dfrac{a\mu}{2}-
\gammae-\dfrac{\pi^2}{9}+\tfrac{2}{3}\right)\cr
+&\dfrac{3c^{(1)}}{2}-2C_F\Bigg\}\dfrac{\alpha_s}{\pi};
\cr}\equn{(5.1)}$$
$c^{(1)}$ given in (2.1) and the nonperturbative contribution is  
$$
\delta_{\rm NP}=\tfrac{1}{2}
\left[\tfrac{270\,459}{108\,800}+\tfrac{1\,838\,781}{2\,890\,000}\right]
\dfrac{\pi\langle\alpha_s G^2\rangle}{m^4\widetilde{\alpha}_s^6}.$$
 
The corrections  are {\sl very} large. Because of this the calculation is
unreliable, even for $\bar{b}b$, 
and fails completely for $\bar{c}c$. 
One might hope that this would be arranged by the
$O(\alpha_s^2)$ corrections (NNLO); but this appears not to be the case. 
We have, first, ``hard" corrections\ref{26}, 
similar to $\delta_{\rm rad}$; 
and corrections to the wave function, given in 
refs.~11,27 where we send for the (rather lengthy) 
explicit formulas. 

Numerically, and with the width in \kev,

\bigskip
\setbox1=\vbox{\offinterlineskip\hrule
\halign{
&\vrule#&\strut\hfil#\hfil&\vrule#&\strut\quad#\quad&\vrule#&
\strut\quad#\quad&\vrule#&\strut\quad#\quad&\vrule#&\strut\quad#&#\cr
 height2mm&\omit&&\omit&&\omit&&\omit&&\omit&\cr 
&\kern0.5mm$\Gammav(\Upsilonv\to e^+e^-)$\kern0.5mm&
&\hfil LO\hfil&&\hfil NLO\hfil&&\hfil NNLO\hfil& \cr
 height1mm&\omit&&\omit&&\omit&&\omit&\cr
\noalign{\hrule} 
height1mm&\omit&&\omit&&\omit&&\omit&\cr
&$\mu=m$&& $0.41$&&$1.22$\vphantom{$4^{4^4}_{4_4}$}&&
$1.13$&\cr
&$\mu=2/a$&& $0.73$&&$0.55$\vphantom{$4^{4^4}_{4_4}$}&&
$0.80$&\cr
 height1mm&\omit&&\omit&&\omit&&\omit&\cr
\noalign{\hrule}}
\vskip.05cm}
\centerline{\box1}
\smallskip

\noindent The experimental figure is
$$\Gammav(\Upsilonv\to e^+e^-)=1.32\pm0.04\,\kev.$$
Clearly,  the large  NLO and NNLO perturbative corrections, 
both of similar size, and 
of  the leading NP correction,   
 make the theoretical result unstable, as the Table shows clearly.
 
For the (perhaps measurable) 
toponium (T) width we  get, for $m_t=175\,\gev$ and  
still in \kev,

\bigskip
\setbox1=\vbox{\offinterlineskip\hrule
\halign{
&\vrule#&\strut\hfil#\hfil&\vrule#&\strut\quad#\quad&\vrule#&
\strut\quad#\quad&\vrule#&\strut\quad#\quad&\vrule#&\strut\quad#&#\cr
 height2mm&\omit&&\omit&&\omit&&\omit&&\omit&\cr 
&\kern0.5mm$\Gammav(T\to e^+e^-)$\kern0.5mm&
&\hfil LO\hfil&&\hfil NLO\hfil&&\hfil NNLO\hfil& \cr
 height1mm&\omit&&\omit&&\omit&&\omit&\cr
\noalign{\hrule} 
height1mm&\omit&&\omit&&\omit&&\omit&\cr
&$\mu=m_t$&& $6.86$&&$10.53$\vphantom{$4^{4^4}_{4_4}$}&&
$13.0$&\cr
&$\mu=2/a$&& $10.24$&&$10.91$\vphantom{$4^{4^4}_{4_4}$}&&
$13.5$&\cr
 height1mm&\omit&&\omit&&\omit&&\omit&\cr
\noalign{\hrule}}
\vskip.05cm}
\centerline{\box1}
\smallskip

The situation has improved with respect to what we had 
for bottomium, but there is still a noticeable dependence on the renormalization point, and on the 
order of perturbation theory considered. 
We  conclude on an {\sl estimate} of some 
$11\,-\,14$ \kev\ for the width.
\medskip
 
\noindent{\fib ACKNOWLEDGEMENTS}
\medskip 
\noindent The author is grateful to CICYT, Spain, for partial financial support. 

I would  also like to acknowledge discussions with A. Pineda.

\noindent{\fib REFERENCES}
\smallskip
{\petit
\item{1.}{F. J. Yndur\'ain, ``Heavy Quarkonium",  Lectures
 at the XVII International School of Physics "QCD:
Perturbative or Nonperturbative", Lisbon, 
1999; FTUAM 99-32 (hep-ph/9910399).}
\item{2. }{W. Fischler, Nucl. Phys. {\bf B129} (1977) 157; 
A. Billoire, Phys. Lett. {\bf B92} (1980) 343, to one loop. 
M. Peter, Phys. Rev. Lett. {\bf 78} (1997) 602, with corrections in  
Y. Schr\"oder, Phys. Lett. {\bf B447} (1999) 321, to two loops.}
\item{3.}{W. Buchm\"uller, Y. J. Ng and S.-H. H. Tye, Phys. Rev. {\bf D24} 
(1981) 3003.}
\item{4. }{S. N. Gupta and S. Radford, Phys. Rev. {\bf D24} (1981) 2309 and (E) 
{\bf D25} (1982) 3430; S. N. Gupta, S. F. Radford 
and W. W. Repko, {\sl ibid} {\bf D26} (1982) 3305.}
\item{5. }{S. Titard and F. J. Yndur\'ain, Phys. Rev. {\bf D49} (1994) 6007; 
 Phys. Rev. {\bf D51} (1995) 6348.}
\item{6. }{N.~Brambilla et al.,  Phys. Lett. {\bf B470} (1999) 215.}
\item{7. }{W. E. Caswell and G. P. Lepage,  Phys. Lett. {\bf B167} (1986) 437.}
\item{8. }{A. Pineda ad J. Soto, Nucl. Phys. B (Proc. Suppl.) {\bf 64} (1998) 428.}
\item{9. }{A. Pineda and F. J. Yndur\'ain,  Phys. Rev. {\bf D58} (1998) 094022, 
and {\bf D61}, 077505 (2000).}
\item{10. }{N.~Brambilla et al., hep-ph/9903355, in press in Phys. Rev. 
The study of the singularities of the static QCD potential 
was first carried out by T. ~Appelquist, M.~Dine and
 I.~J.~Muzinich, Phys. Rev. {\bf D17} (1978) 2074.}
\item{11. }{F. J. Yndur\'ain, hep-ph/0002237 (2000);  
 FTUAM 00-14, hep-ph/0007333 (Proc. E. Schr\"odinger Institute Seminar on Confinement, 
May -- June, 20000, 
to be published).}
\item{12. }{J.~Santiago and F. J. Yndur\'ain, 
Nucl. Phys. {\bf B563} (1999) 45, and references therein}
\item{13. }{W. Lucha and F. F. Sch\"oberl, UWThPh-1999-77 (hep-ph/0001191)}
\item{14. }{R. Coquereaux, Phys. Rev. {\bf D 23} (1981) 1365; R.~Tarrach, 
 Nucl. Phys.  {\bf B183} (1981) 384; to two loops, 
N.~Gray et al., Z. Phys.  {\bf 48} (1990) 673. 
The $O(\alpha_s^3)$ corrections to the relations between 
the pole-\msbar\ masses have been recently evaluated by 
K.~Melnikov and T.~van~Ritbergen, hep-ph/9912391.}
\item{15. }{M. Beneke and A. Signer, Phys. Lett. {\bf B471} (1999) 233.}
\item{16. }{N. Brambilla and A. Vairo, CERN-TH/20000-036 (hep-ph/0002075).}
\item{17. }{M. B. Voloshin, Nucl. Phys. {\bf B154} (1979) 365 and Sov. J. Nucl. Phys. 
{\bf 36} (1982) 143; H. Leutwyler, Phys. Lett. {\bf B98} (1981) 447.} 
\item{18. }{ H. Dosch, Phys. Lett. {\bf B190} (1987) 177; Yu. A. Simonov, 
Nucl. Phys. {\bf B307} (1988) 512 and {\bf B324} (1989) 56; H. Dosch and 
 Yu. A. Simonov, Phys. Lett. {\bf B205} (1988) 339; 
Yu. A. Simonov, S. Titard and  F. J. Yndur\'ain, Phys. Lett. {\bf B354} (1995) 435.}
\item{19. }{U. Aglietti and Z. Ligeti, Phys. Lett. {\bf B364} (1995) 75.}
\item{20. }{R. Akhoury and V. I. Zakharov,  
Phys. Lett. {\bf B438} (1998) 165; F.~J.~Yndur\'ain, 
 Nucl. Phys. B (Proc. Suppl.) {\bf 64} (1998) 433.}
\item{21. }{I. I. Bigi and N. G. Uraltsev, Phys. Lett. {B321} (1994) 412; 
I. I. Bigi  et al., Phys. Rev. {\bf D50} (1994) 2234.}
\item{22. }{A. Pineda, Ph. D. Thesis, Univ. Barcelona (1998), 
A.~Hoang, M.~Smith, T. Stelzer and S. Willenbrock, Phys. Rev. {\bf D59} (1999) 114014
and 
M.~Beneke, Phys.Lett. {\bf B434} (1998) 115 
for the constant term; and  
N. Brambilla et al., {\it Phys. Rev.} D {\bf 60} (1999) 091502  and 
Nucl. Phys. {\bf B566} (2000) 275  for the term in $C_1$ in (3.3).}
\item{23. }{Yu. A. Simonov, ``QCD and Topics in Hadron Physics", Lectures
 at the XVII Int'nal School of Physics "QCD:
Perturbative or Nonperturbative", Lisbon, 
1999, hep-ph/9911237.}
\item{24. }{J. L. Richardson, Phys. Lett. {\bf B82} (1979) 272. 
For an application to deep inelastic scattering, 
see K. Adel, F.~Barreiro and F. J. Yndur\'ain, Nucl. Phys.,
 {\bf B495} (1997) 221.}
\item{25. }{R. Barbieri et al., Phys. Lett. {\bf 57B} (1975) 455; {\sl ibid.} Nucl. 
Phys. {\bf B154} (1979) 535.}
\item{26. }{A. H. Hoang and T. Teubner, 
Phys. Rev. {\bf D58} (1998) 114023; 
A.~Czarnecki and K.~Melnikov, 
Phys. Rev. Lett. {\bf 80} (1998) 2531; M.~Beneke, A.~Singer and 
V.~A.~Smirnov, Phys. Rev. Lett. {\bf 80} (1998) 2535.}
\item{27. }{A.~A. Penin and A.~A.~Pivovarov, Nucl. Phys. {\bf B549} (1999) 217;
K. Melnikov and A. Yelkhovsky, Phys. Rev. {\bf D59} (1999) 114009.}
\item{}{}
}
}\enddc
\bye